      [1999/12/01 v1.4c Il Nuovo Cimento]
\documentclass{cimento}

\usepackage{graphicx}
\usepackage{epstopdf}

\title{Full one-loop electroweak corrections to $e^+e^-\to 3$
jets \ETC \ at linear colliders}
\author{C.M.~Carloni~Calame \from{ins:a}\ETC,
S.~Moretti \from{ins:a},
F.~Piccinini \from{ins:b} \atque
D.A.~Ross \from{ins:a}}
\instlist{\inst{ins:a} School of Physics and Astronomy, University of
Southampton, Southampton, UK
  \inst{ins:b} Istituto Nazionale di Fisica Nucleare, Sezione di
  Pavia, Pavia, Italy}
\PACSes{\PACSit{13.66.Bc}{Hadron production in $e^+e^-$ interactions}
\PACSit{12.15.Lk}{Electroweak radiative corrections}
\PACSit{14.70.Hp}{Z bosons}}

\begin{document}
\maketitle
\begin{abstract}
We describe the impact of the full one-loop electroweak terms of
${\cal O}(\alpha_{\mathrm{S}} \alpha_{\mathrm{EM}}^3)$ entering the electron-positron into
three-jet cross-section from $\sqrt{s}=M_Z$ to TeV scale energies.
We include both factorisable and non-factorisable virtual corrections
and photon bremsstrahlung.
Their importance for the measurement of $\alpha_{\mathrm{S}}$ from jet rates and 
shape variables is explained qualitatively and illustrated
quantitatively, also in presence of $b$-tagging.
\end{abstract}

\section{Introduction}
\label{introduction}
\noindent
A peculiar feature distinguishing strong (QCD) and electroweak (EW)
effects in higher orders
is that the latter are enhanced by (Sudakov) double logarithmic factors,
$\ln^2(\frac{s}{M^2_{{W}}})$, 
which, unlike in the former, do not cancel for `infrared-safe' 
observables  \cite{Kuroda:1991wn,Beenakker:1993tt,Ciafaloni:1999xg,Denner:2000jv}. 
The origin of these `double logs' is well understood.
It is due to a lack of the Kinoshita-Lee-Nauenberg (KLN) 
\cite{KLN} type 
cancellations of Infra-Red (IR) -- both soft
and collinear -- virtual and real emission in
higher order contributions originating from $W^\pm$ (and, possibly,
$Z$) exchange. 
This is in turn a consequence of the 
violation of the Bloch-Nordsieck theorem \cite{BN} in non-Abelian theories
\cite{Ciafaloni:2000df}.
The problem is in principle present also in QCD. In practice, however, 
it has no observable consequences, because of the final averaging of the 
colour degrees of freedom of partons.
This does not occur in the EW case,
where the initial state has a non-Abelian charge,
dictated by the given collider beam configuration, such as in $e^+e^-$
collisions. 

These logarithmic corrections are finite (unlike in
QCD), as the masses of the weak gauge bosons provide a physical
cut-off for $W^\pm$ and $Z$ emission. Hence, for typical experimental
resolutions, softly and collinearly emitted weak bosons need not be included
in the production cross-section and one can restrict oneself to the 
calculation of weak effects originating from virtual corrections and
affecting a purely hadronic final state.
Besides, these contributions can  be
isolated in a gauge-invariant manner from electromagnetic (EM) effects
\cite{Ciafaloni:1999xg}, 
at least in some specific cases, and 
therefore may or may not
be included in the calculation, depending on the observable being studied. 
As for purely EM effects,
since
 the (infinite) IR real photon emission cannot be resolved experimentally, 
this ought to be combined with the (also infinite) virtual one, through the 
same order, to recover a finite result, which is however not
doubly logarithmically enhanced (as QED is an Abelian theory).
 
In view of all this,  it becomes of crucial importance to assess
the quantitative relevance of such EW corrections
affecting, in particular, key QCD processes studied at past, present and 
future 
colliders, such as $e^+e^-\to3$~jets. 

\section{Calculation}
\label{calculation}
\noindent
In Ref.~\cite{ee3jets}, we calculated the full 
one-loop EW effects entering three-jet production in $e^+e^-$
annihilation at any collider energy
via the subprocesses $e^+e^-\to\gamma^*,Z\to \bar 
qqg$. Ref.~\cite{oldpapers} tackled part of these, restricted to the
case of $W^\pm$ and $Z$ (but not $\gamma$) exchange and 
when the higher order effects arise only from initial or final state
interactions 
(the so-called `factorisable' corrections).
The remainder, `non-factorisable' corrections,
while being typically small at $\sqrt s=M_{Z}$, 
are expected to play a quantitatively relevant role as $\sqrt s$ grows
larger.
We improved on the
results of Ref.~\cite{oldpapers} in two respects: (i) we include now all
the non-factorisable terms; (ii) we also incorporate previously
neglected genuine QED corrections, including photon bremsstrahlung.

A more complete account of the corrections discussed here has recently
appeared in Ref.~\cite{ee3jetsgermans}.

Combining the enhancement associated with the weak Sudakov logarithms
to the decrease of $\alpha_{\mathrm{S}}$ with energy, 
in general, one expects one-loop EW effects to become comparable to QCD ones
at future Linear Colliders (LCs) \cite{LCs} running at TeV energy scales,
like those available at an International Linear Collider (ILC) or the
Compact LInear Collider (CLIC).
In contrast, at the $Z$ mass peak, where logarithmic enhancements are
not effective, one-loop EW corrections are expected to appear
at the percent level, hence being of limited relevance at
LEP1 and SLC, where the final error on $\alpha_{\mathrm{S}}$
is of the same order or larger,
but of crucial importance at a GigaZ stage of a future
 LC \cite{oldpapers}, where the relative accuracy
of $\alpha_{\mathrm{S}}$ measurements is expected to be at the
$0.1\%$ level or better.
Concerning higher order QCD
effects, a great deal of effort has    
recently been devoted to evaluate two-loop contributions
to the three-jet process \cite{QCD2Loops}
while the one-loop QCD results have been known for quite some time \cite{ERT}.

In $e^+e^-$ annihilations, the most important QCD quantity to be 
extracted from multi-jet events is $\alpha_{\mathrm{S}}$.
The confrontation of the measured value of the strong coupling
constant with that predicted by the theory through the 
renormalisation group evolution is an important test of the Standard Model
(SM). Alternatively, it may be an indication of new physics, when its 
typical mass scale is larger than the collider energy, so that 
the new particles cannot be
produced as `real' detectable states
but may manifest themselves through `virtual' effects. 
Not only jet rates,
but also
jet shape observables would be affected.

\begin{figure}
\begin{center}
\includegraphics[width=10cm]{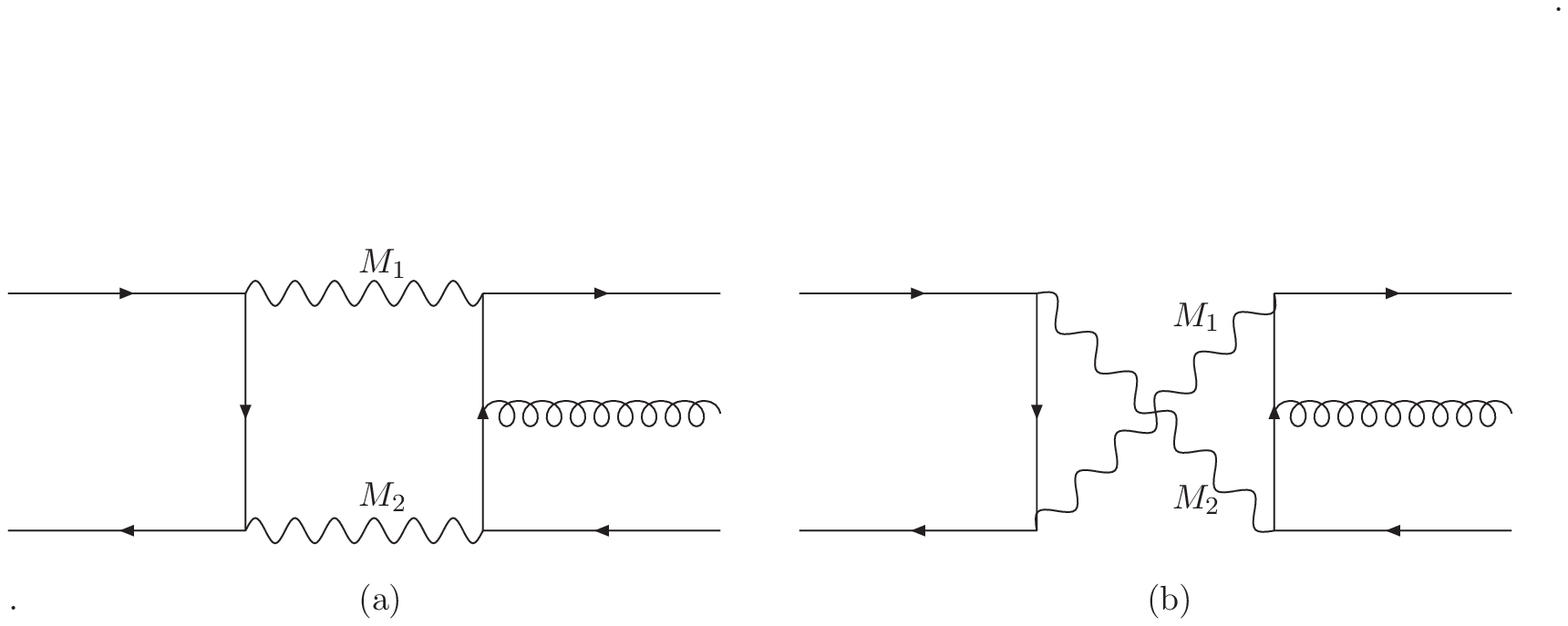}
\end{center}
\caption{Pentagon graphs}
\label{fig:pentagons}
\end{figure}

The detailed discussion of the calculation can be found in
Ref.~\cite{ee3jets}. Here, for the sake of completeness, we mention
that the calculation of virtual corrections is performed in the
't~Hooft-Feynmann gauge.
It is also worth mentioning that initial state
electron-positron polarisations are retained and it is possible to
study EW effects in presence of polarised incoming beams. For genuinely weak interaction
corrections, this is of particular interest, since such corrections violate 
parity conservation. 

In the calculation IR divergences are regulated by means of a small
photon mass $\lambda$, both in virtual and real QED corrections. The
independence of the final results from the photon mass has been successfully checked.

A new feature of this calculation
is the
occurrence of pentagon graphs, as those shown in Fig.~\ref{fig:pentagons}.
We have handled these in two 
separate ways (with two independently developed codes), in order to check
for possible numerical instabilities, finding good agreement.

The collinear QED divergence
gives rise to a large logarithm ($\ln(s/m_f^2)$),
which is associated with the Initial State Radiation (ISR) induced by
the incoming electrons and positrons.
In the case of
electron-positron colliders this large correction is always present
and it is universal to all processes.
For sensible numerical 
results, it has to be accounted for to all orders of perturbation 
theory, e.g., within the so-called electron/positron structure function formalism~\cite{sf}, 
which automatically resums in QED all Leading Logarithmic (LL) terms. 
In Ref.~\cite{MNP} a method of combining consistently 
resummed LL calculations with exact ${\cal O}(\alpha_{\mathrm{EM}})$ ones 
has been devised both in additive and factorisable form. 
Here, we adopted the additive approach.

In order to integrate over the phase-space,
the width, $\Gamma_Z$, of the $Z$ boson has been included in the propagator. 
For consistency, this means that the same width has to 
be included in the $Z$ propagator for the virtual corrections. 
The essential ingredient for the 
evaluation of virtual corrections is the ability to compute 
one-loop integrals with complex internal masses.
We implemented the general expression 
for the scalar four-point function of Ref.~\cite{tHV}, 
valid also for complex masses.
Particular attention has been 
devoted to the occurrence of numerical instabilities in 
certain regions of phase space because of strong cancellations. 

We have neglected the masses of light quarks throughout. However, 
in the case in which the final state
contains a $b\bar b$ pair, whenever there is a $W^\pm$ boson
in the virtual loops, account had to be taken of the mass of the 
top (anti)quark.
We are therefore in a position to present the results for such 
`$b$-jets' separately, as reported in~\cite{eebjets}.

\section{Numerical results}
\label{numbers}
\noindent
The numerical results presented in this section are obtained
considering a realistic experimental setup. The input parameters and
the setup of the cuts is described in Ref.~\cite{ee3jets}. A Cambridge
jet algorithm is used to cluster parton momenta into jets. 
Finally, we sum over the final-state quarks, if not stated otherwise.

\begin{figure}\begin{center}{
\includegraphics[width=6cm]{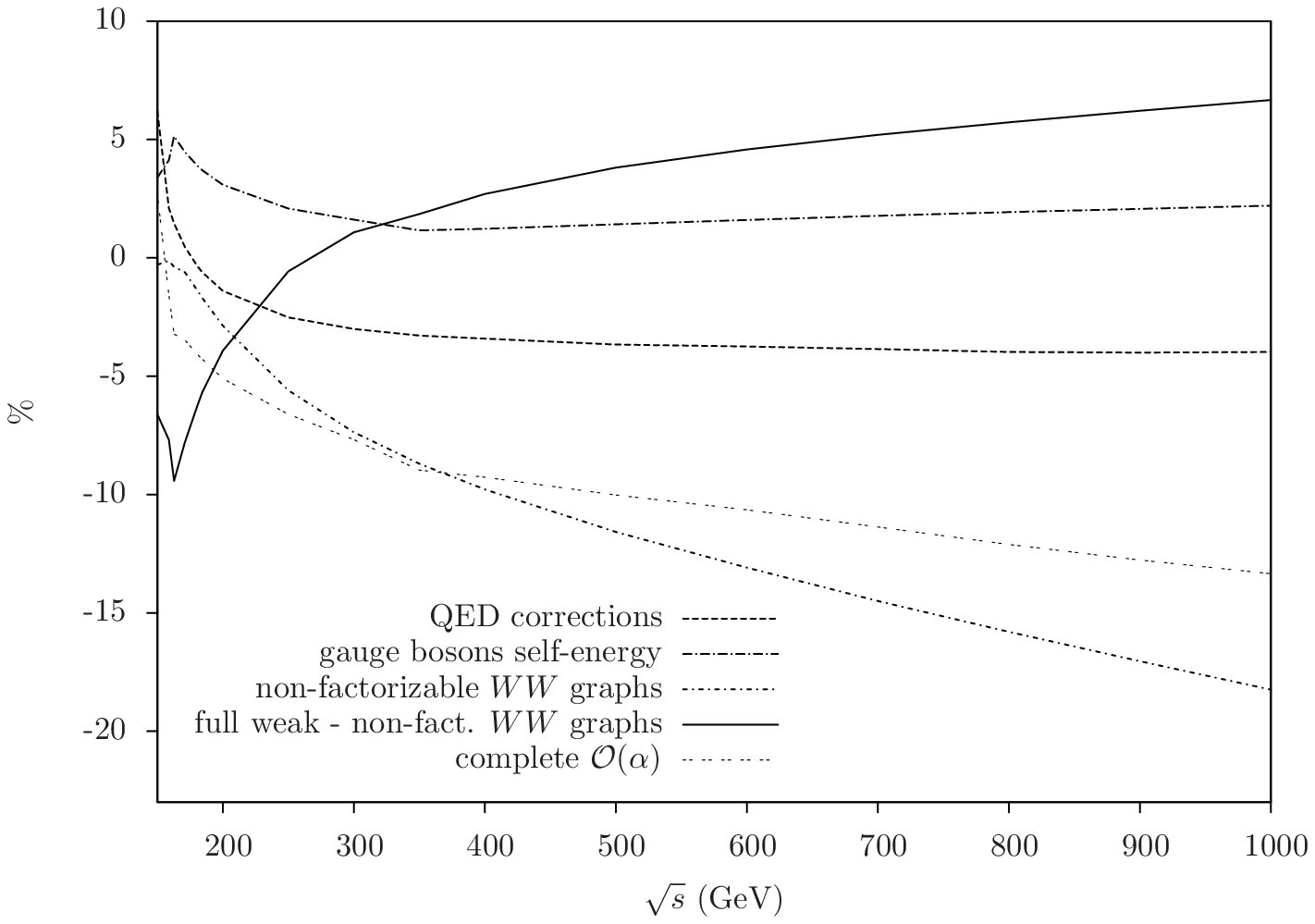}~\includegraphics[width=6cm]{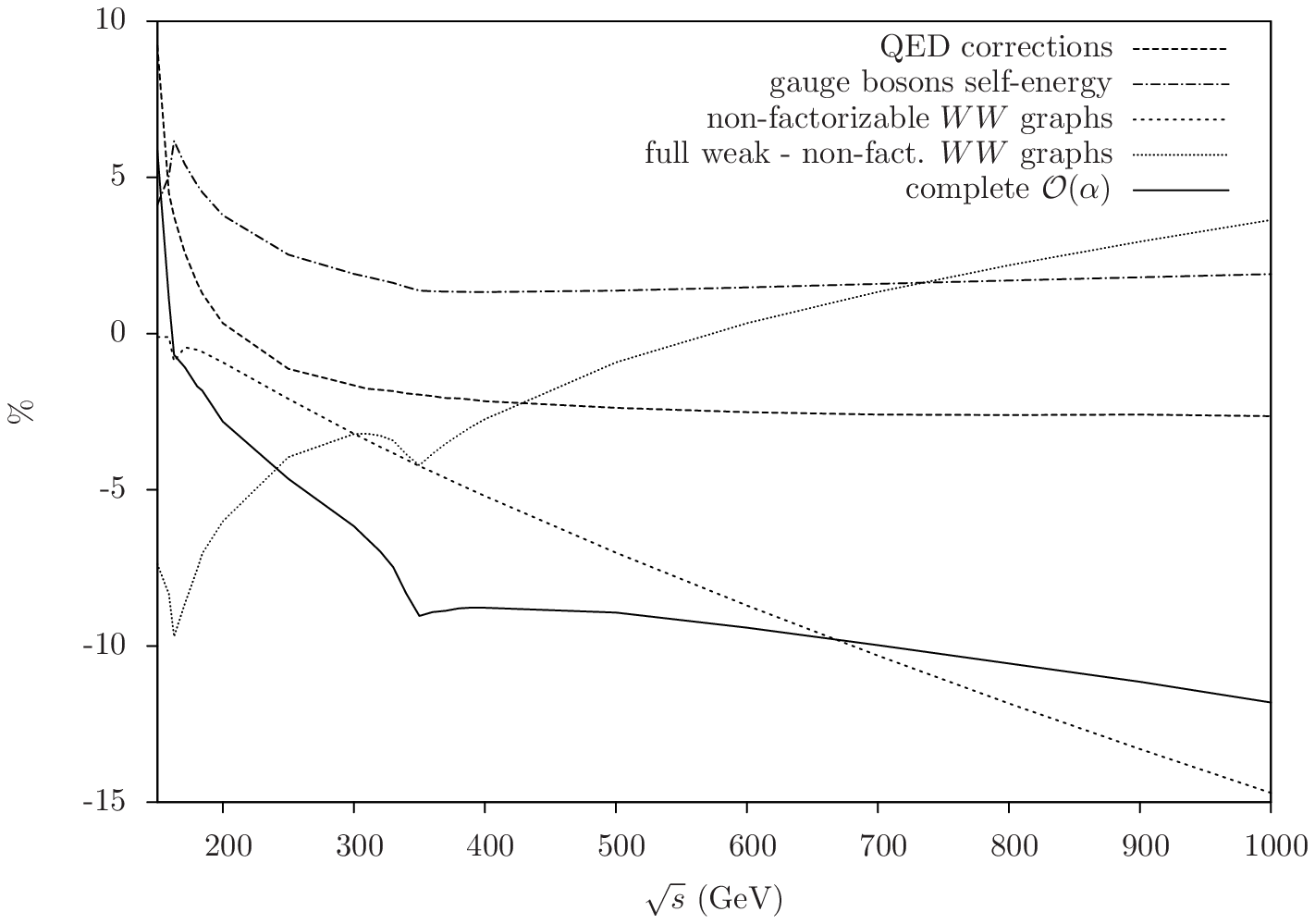}
\caption{Relative effect on the integrated cross section due to
  different contributions to the order $\alpha\equiv
\alpha_{\mathrm{EM}}$ correction, as a function of the CM energy.
On the left the sample inclusive over the quark flavours is shown, on the right
the $b$-jets subsample is considered.
}
\label{energyscan}}
\end{center}
\end{figure}
On the left of Fig.~\ref{energyscan}, the relative effects
 on the cross
section 
induced by different contributions to the order $\alpha_{\mathrm{S}}
\alpha_{\mathrm{EM}}^3$
correction are plotted as a function of the CM energy, in
the range from 150 GeV to 1 TeV, when considering a sample summed over
the quark flavours. The curves represent the effect of
the QED
corrections only, the effect of the gauge
bosons self-energy corrections,
the
effect of the non-factorisable graphs
with $WW$
exchange,
the effect of
the weak corrections with the non-factorizing $WW$ graphs removed
(labelled as ``full weak - non-fact $WW$ graphs'') 
 and the total effect as the sum of the previous
ones: the total effect is increasingly negative, reaching the $-13\%$ level at
1 TeV. 
It is worth mentioning that, as far as the non-factorisable
$WW$ corrections 
are concerned, in the case of $d$,
$s$ and $b$ final-state quarks, only the direct diagrams
are present due to charge conservation, while, for $u$ and $c$ quarks, only
crossed diagrams are present, if the sum over initial- and final-state
helicities is taken. In the case of $ZZ$ exchange, all the graphs survive,
giving rise to a cancellation at the leading-log level between direct
and crossed diagrams, which does not occurr for $WW$ exchange. Hence,
the big negative correction is due to the presence of the
$WW$ non-factorisable graphs, which develop the aforementioned 
large Sudakov double logarithms in the high energy regime. In the
right panel of Fig.~\ref{energyscan}, the corrections to the process
$e^+e^-\to b\bar{b}g$ are shown, assuming that an efficient
$b$-tagging is present.

We then show the impact of the EW corrections on some
differential distributions of phenomenological interest.
The plots show the tree-level contributions and the higher order
corrections in three different 
contributions: the purely weak-interaction contribution 
(labelled ``weak ${\cal O}(\alpha)$"), purely weak plus QED corrections,
which are dominated by the above-mentioned ISR (labelled ``exact ${\cal
O}(\alpha)$"), and the weak plus electromagnetic 
correction in which the LL have been summed (labelled ``exact
${\cal O}(\alpha)$ + h.o. LL"). The figures show in the upper panel
the absolute distributions and in the lower panel the relative differences with
respect to the tree-level rates.

\begin{figure}
\begin{center}
\includegraphics[width=6cm]{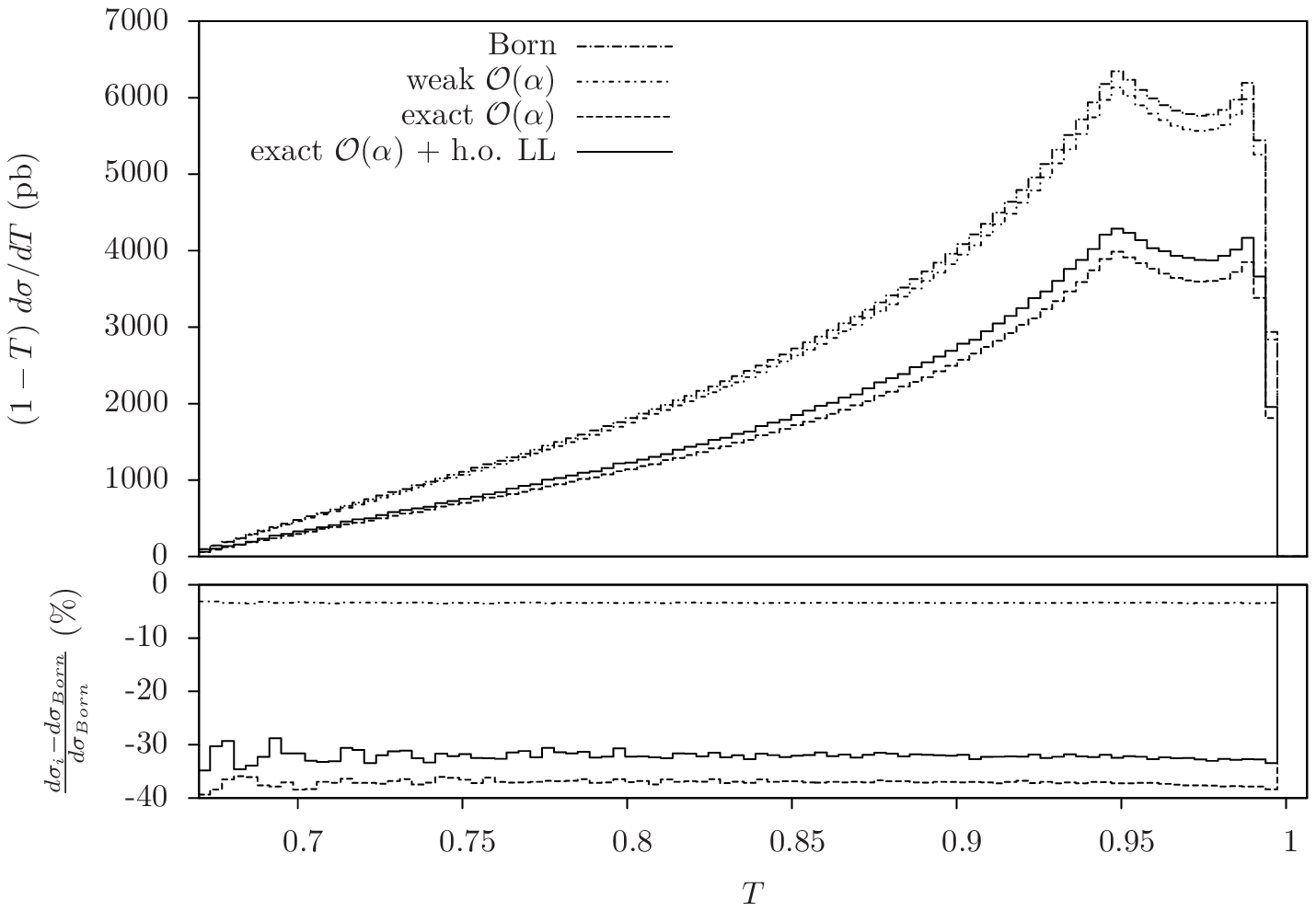}~\includegraphics[width=6cm]{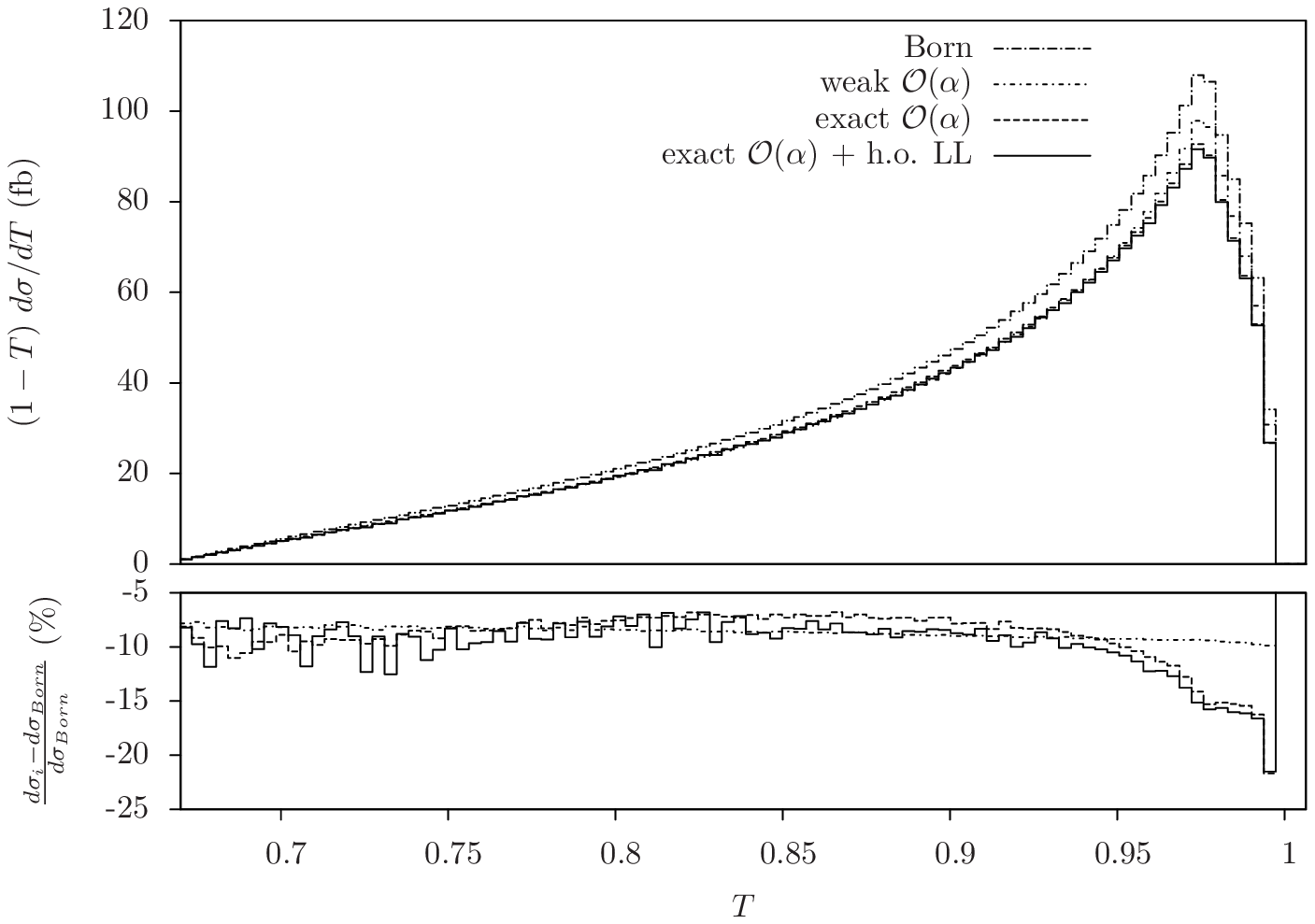}
\caption{$(1-T)\frac{d\sigma}{dT}$ distribution at the $Z$ peak (left)
and at 1 TeV (right).}
\label{thrust-1000}
\end{center}\end{figure}
In Fig.~\ref{thrust-1000}, the {\em thrust} event shape distribution
is shown, in the form $(1-T)\frac{d\sigma}{dT}$.
The $T$ distribution is one of the key observables used for the
measurement of $\alpha_{\mathrm{S}}$ in $e^+e^-$
collisions~\cite{kunsztnason}. It is worth noticing that while the
purely weak corrections give an almost constant effect on the whole
$T$ range, the presence of the real bremsstrahlung gives a non trivial
effect in the region $T>0.92$. In view of a precise measurement of
$\alpha_{\mathrm{S}}$ at a future LC, EW corrections can
play an important role.

\begin{figure}\begin{center}
\includegraphics[width=6cm]{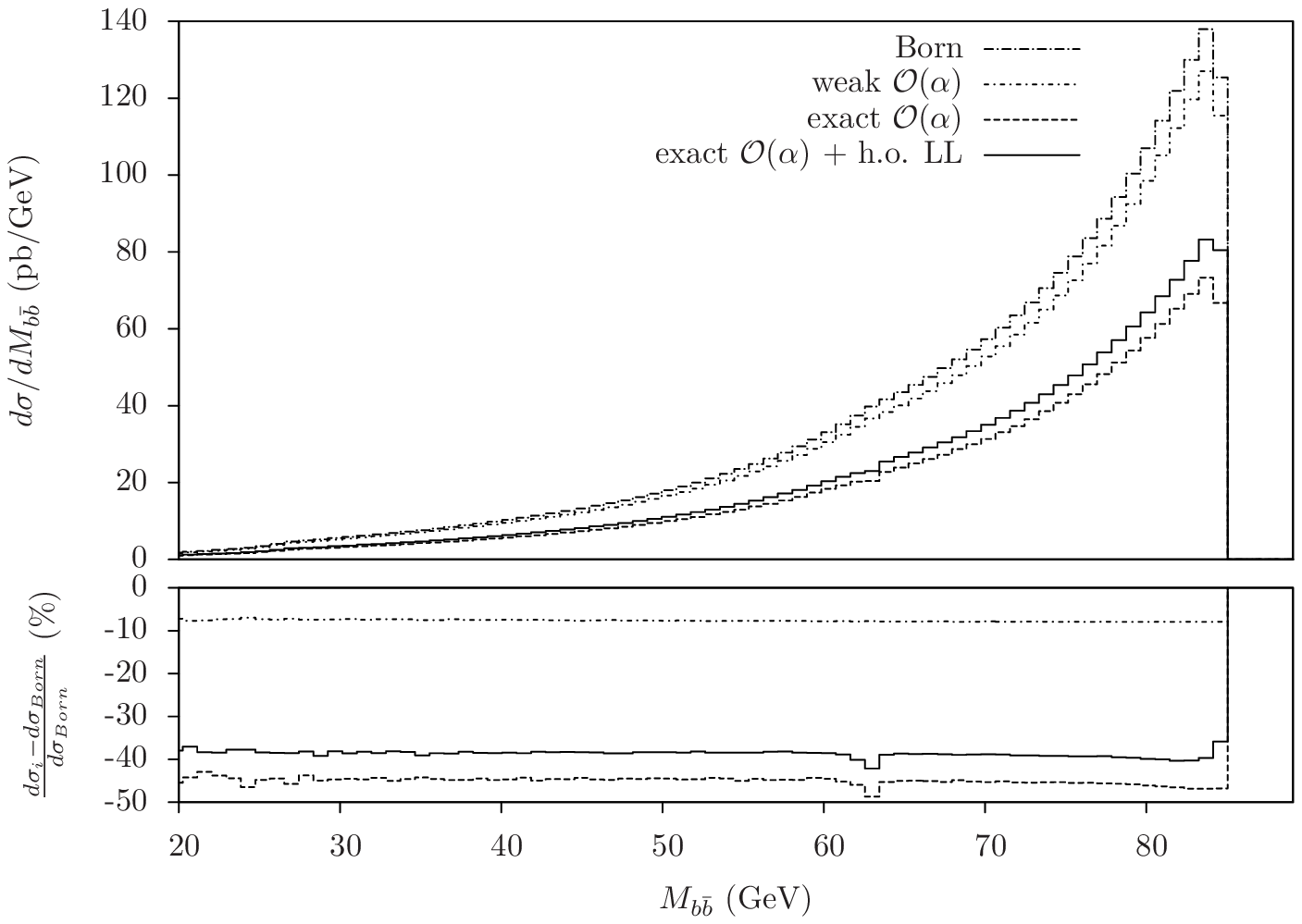}~\includegraphics[width=6cm]{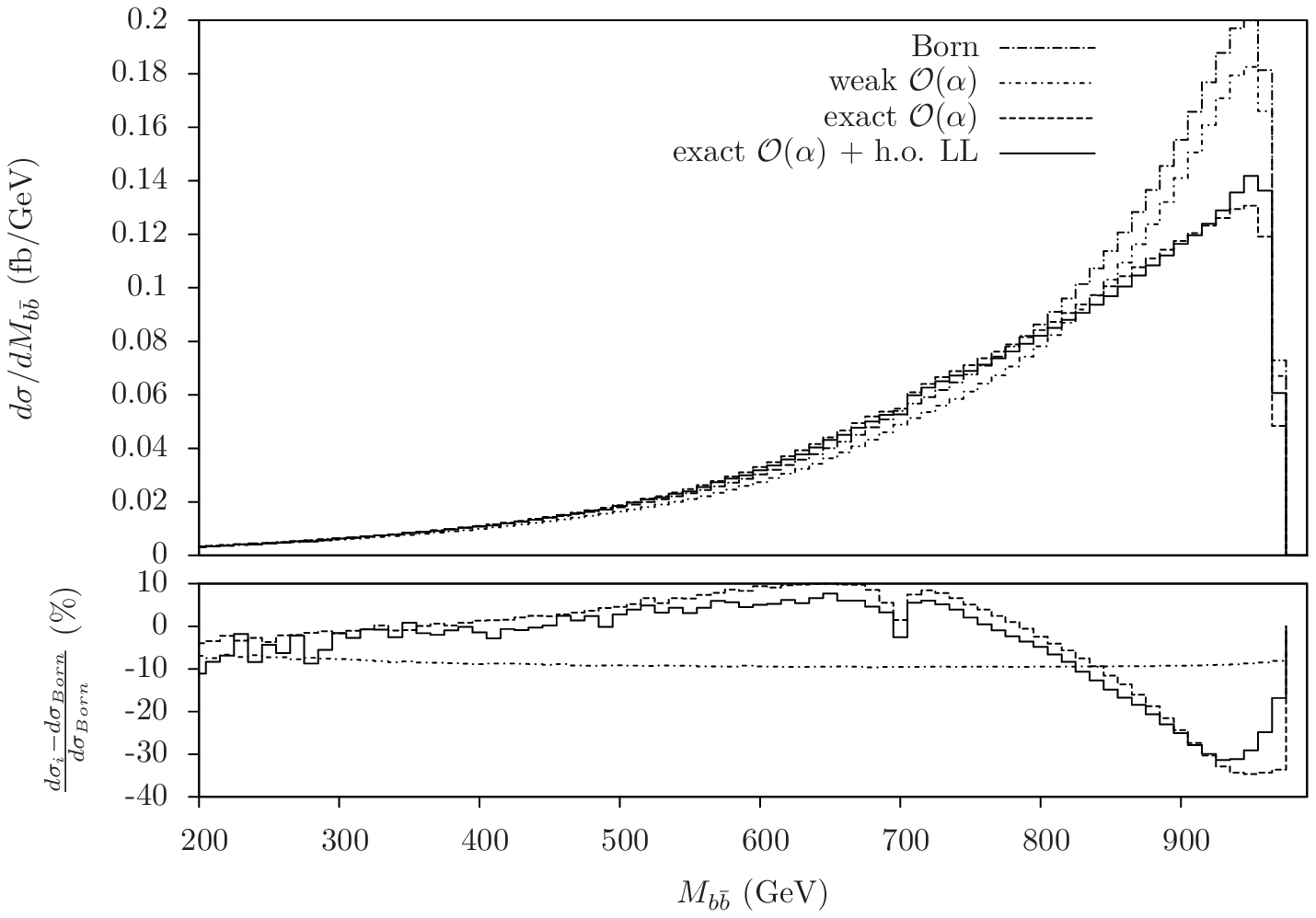}
\caption{$b\bar b$ invariant mass distribution at the $Z$ peak (left)
and at 1 TeV (right).}
\label{minbbj-peak}\end{center}\end{figure}

The ability to efficiently tag $b$-quark jets enables one to define
observables in $b\bar bg$ final states
which are not (easily) reconstructable in the case
of the full three-jet sample. One example is the invariant mass
of the $b\bar b$ pair, $M_{b\bar b}$, which we plot in 
Fig.~\ref{minbbj-peak}.
Here, the largest contribution to the total correction  comes
from QED ISR, primarily because of the radiative return phenomenon.

\section{Conclusions}
\noindent
In summary, we have shown the phenomenological relevance that the calculation
up to
${\cal O}(\alpha_{\mathrm{S}}\alpha_{\mathrm{EM}}^3)$
can have in the study of (unflavoured) three-jet samples
in $e^+e^-$ annihilation, for all energies ranging from $\sqrt s=M_Z$ to
1 TeV. Not only inclusive jet rates are affected, but also more exclusive
distributions, both global (like the event shape variables) 
and individual (like invariant mass) ones. Effects range from a few percent to several tens
of percent, depending on the energy and the observable being studied,
and we have shown cases where such higher-order contributions would impinge
on the experimental measurements of jet quantities.
Finally, notice that, depending on experimental procedures, a
different normalisation of the distributions, 
like, e.g., the one adopted in Ref.~\cite{ee3jetsgermans}, would lead
to somewhat different corrections in general.

\acknowledgments
\noindent
C.M. Carloni Calame would like to warmly thank the Local Organizing Committee
for the pleasant and stimulating atmosphere during the LC09 conference.

\end{document}